# ECGformer: Leveraging transformer for ECG heartbeat arrhythmia classification

Taymaz Akan
*Department of Medicine*
*Louisiana State University Health Sciences Center*
Shreveport, USA
taymaz.farshi@gmail.com

Sait Alp
*Department of Computer Engineering*
*Erzurum Technical University*
Erzurum, Turkey
sait.alp@erzurum.edu.tr

Mohammad Alfrad Nobel Bhuiyan
*Department of Medicine*
*Louisiana State University Health Sciences Center*
Shreveport, USA
nobel.bhuiyan@lsuhs.edu

*Abstract—* **An arrhythmia, also known as a dysrhythmia, refers to an irregular heartbeat. There are various types of arrhythmias that can originate from different areas of the heart, resulting in either a rapid, slow, or irregular heartbeat. An electrocardiogram (ECG) is a vital diagnostic tool used to detect heart irregularities and abnormalities, allowing experts to analyze the heart's electrical signals to identify intricate patterns and deviations from the norm. Over the past few decades, numerous studies have been conducted to develop automated methods for classifying heartbeats based on ECG data. In recent years, deep learning has demonstrated exceptional capabilities in tackling various medical challenges, particularly with transformers as a model architecture for sequence processing. By leveraging the transformers, we developed the ECGformer model for the classification of various arrhythmias present in electrocardiogram data. We assessed the suggested approach using the MIT-BIH and PTB datasets. ECG heartbeat arrhythmia classification results show that the proposed method is highly effective.**

## I. INTRODUCTION

cardiovascular diseases (CVDs) are a group of conditions that hurt the cardiovascular system, which includes the heart and blood vessels. CVDs are consistently ranked highly among the leading causes of death across the globe [1]. They can be structural problems like coronary artery disease, heart failure, and birth defects of the heart, or they can be functional problems like arrhythmias [2]. An arrhythmia is a condition where the heart beats in an abnormal rhythm, either too fast, too slow, or irregularly. It is classified as a cardiovascular condition due to its impact on the heart's ability to efficiently pump blood, which can result in complications and affect overall cardiovascular well-being [3]. Arrhythmias come in different forms and can be triggered by a range of factors, such as heart disease, imbalances in electrolytes, and other medical conditions [4].

Heart rhythm abnormalities can range from lack of symptoms to sudden cardiac arrest, leading to sudden cardiac death (SCD). SCD is a major public health issue, accounting for 50-60% of deaths in patients with coronary artery disease. Survival rates are only 3-10% in hospital settings, highlighting the need for risk prediction, prevention, and adequate treatment of arrhythmias [5], [6]. Common symptoms of cardiac arrhythmias include fluttering, pounding, shortness of breath, chest pain dizziness, palpitations, rapid heart rate, and a sense of weakness [7]. Diagnosing cardiac arrhythmias involves observing blood pressure, ECG readings, and irregular heartbeats, as well as noting symptoms such as weakness, fatigue, dizziness, and reduced activity levels in daily routines. The electric activity of the heart is measured by ECG, which has been extensively utilized in the detection of heart diseases because of its simplicity and non-invasive characteristics [8].

Up until recently, the examination of ECGs was carried out manually by healthcare professionals. However, automatic detection of arrhythmias is crucial for early intervention and treatment. It allows for early identification and treatment, reducing the risk of complications and heart attacks. Continuous monitoring, using wearable devices or implantable monitors, provides a comprehensive picture of cardiac health. Automatic detection systems improve accuracy and efficiency, reducing human error. Remote monitoring and telemedicine make it feasible for healthcare providers to monitor patients remotely. Enhanced patient care promotes awareness and proactive measures for maintaining a healthy lifestyle.

There has been significant attention on the examination of ECG signals for automatic detection of cardiac arrhythmia [9], [10]. Different conventional machine learning methods [8], [11]–[13] have been used to analyze and classify ECG signals, such as multi-layer perceptron, support vector machines, random forests, and decision trees. An ECG can be represented as a time series, also known as sequential data [14]. Sequential or temporal order of data points is not always considered by traditional algorithms. As a result, they may miss important dependencies or patterns in sequential data because they treat each instance as separate. Additionally, they require manual engineering and selection of relevant features, which can be difficult and potentially miss intricate details in complex data. Deep learning models, on the other hand, can learn hierarchical representations from raw data automatically, so feature engineering doesn't have to be done by hand. From the data they are given, they can find complex patterns and representations.

In recent years, deep learning has sparked significant innovation in various domains, including medical applications [15]–[17]. Several advanced deep learning techniques, including belief propagation deep neural networks (DNNs), convolutional neural networks (CNNs) [18], [19], recurrent neural networks (RNNs) [20], [21], and Transformers [22], [23], have been employed to study arrhythmias and analyze ECG signals. Moreover, DNNs and CNNs methods struggle to effectively learn long-term dependencies from long ECG sequences. transformers and RNNs, on the other hand, are capable of learning the long-term dependence of an ECG sequence.

Over the past few years, the use of deep learning techniques has become increasingly popular in the field of ECG classification. These techniques have shown promising results in detecting arrhythmias, as evidenced by several studies [22]–[26]. Transformers offer several advantages over other deep learning models, such as recurrent neural networks (RNNs) and convolutional neural networks (CNNs). They excel in capturing long-range dependencies in sequences, making them more computationally efficient. Transformers process input sequences in parallel, making them more scalable than RNNs. They use attention mechanisms to focus on relevant information, enabling them to handle positional information effectively. They can capture complex relationships between elements in the sequence without relying on predefined structures, making them particularly effective for tasks involving sequential data, natural language processing, and applications requiring understanding long-range dependencies. This paper employs a transformer model with attention mechanisms to classify arrhythmias from ECG time series data.

## II. PROPOSED METHOD

The term "transformer" was initially introduced in the domain of machine translation. The Transformer Neural Network is designed to handle long-range dependencies, avoid recursion, and enable parallel computation, resulting in reduced training time and improved performance [27]. Transformers in natural language processing (NLP) involve an encoder and decoder, converting input into hidden layers and back into natural language sequences [28]. These self-attention-based architectures, focusing on attention mechanisms, have revolutionized deep learning in areas like NLP, computer vision, and audio processing. Their critical features include non-sequential, self-attention, and positional embeddings. The core is composed of a sequence of encoder and decoder layers. In this paper we Leveraged a transformer for ECG heartbeat arrhythmia classification. Inspired by the Transformer scaling successes in NLP, we experiment with applying a standard Transformer directly to ECG signal sequences. To do so, we split a series of waves into patches and provide the sequence of linear embeddings of these patches as an input to a Transformer. signal patches are treated the same way as tokens (words) in an NLP application.

Every encoder consists of two primary sublayers: a multi-head attention layer and a position-wise fully connected FFN (as shown in Fig. 1 (d). In addition to these two sublayers, there are Residue skip connections present around both layers, as well as two LayerNorm layers (see Fig. 2). In the self-attention layer, attention weights are calculated between each time step and every other time step in the time series. The attention weights range from 0 to 1. In the realm of natural phenomena, a remarkable process known as scaled dot-product attention function takes center stage. This process involves the transformation of a query vector $Q$ and a collection of key-value $K$ pairs into a magnificent output vector $V$. The significance of every time step in capturing the temporal patterns of the time series is evident in these attention weights. Here is the equation that is used to calculate the attention weights (output matrix):

$$head = \text{attention}(Q, K, V) = \text{softmax}\left(\frac{QK^\top}{\sqrt{d_k}}\right)V \quad (1)$$

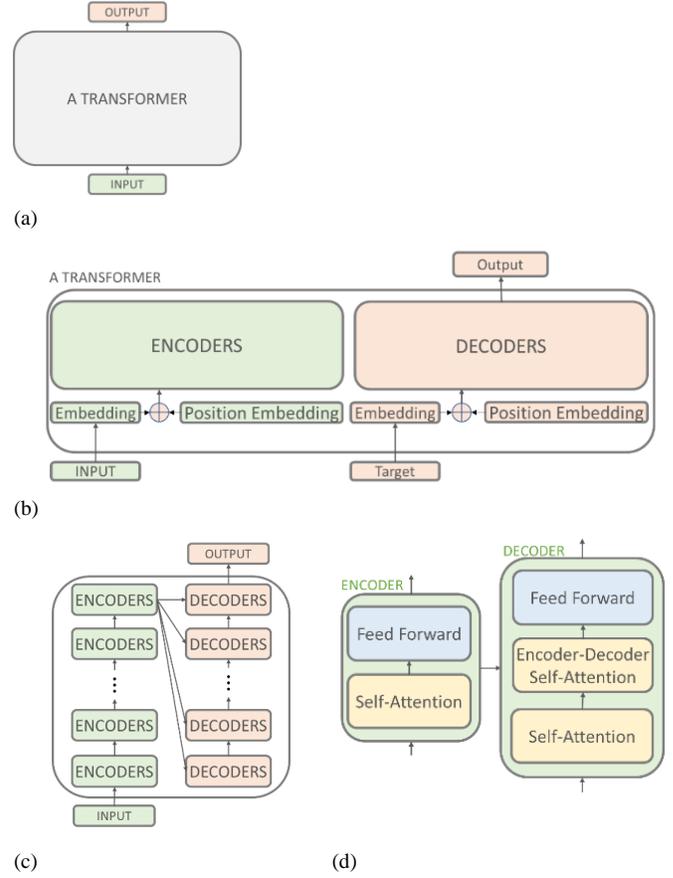

(a)

(b)

(c)                    (d)

Fig. 1. General Overview of Transformers

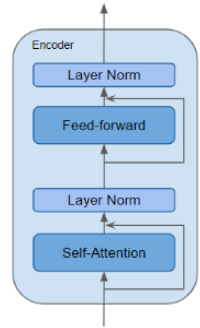

Fig. 2. General Overview of Transformers

The attention weights are normalized using the SoftMax function to ensure that their total sum is equal to 1. The square root of $d_k$ serves as a crucial factor in maintaining gradient stability throughout the training process. In contrast to the sequence-to-sequence transformer, where the Q, K, and V matrices differ, the time series classification transformer utilizes identical matrices for all operations. A multi-head (MA) attention layer consists of multiple parallel scaled dot product attention layers, each known as a head. The results from the H heads are combined and projected onto another dense layer to generate the ultimate hidden representation. The multi-head attention function can be described as follows:

$$\text{multihead}(Q, K, V) = \text{concat}(\text{head}_1, \text{head}_2 \cdots \text{head}_h)W^O \quad (2)$$

$$\text{head}_i = \text{attention}\left(QW_i^Q, KW_i^K, VW_i^V\right) \quad (3)$$

Herein, the weight matrices $QW_i^Q$, $KW_i^K$, and $VW_i^V$ are used to transform the input matrix $X$ into the query, key, and value spaces. Finally, a SoftMax function processes the feed-forward layer's output to create a probability distribution across the R classes. The time series class label is predicted based on the class with the highest probability. It is worth noting that our method has no Decoder at all and rely only on the Encoder.

## III. Dataset

The MIT-BIH Arrhythmia Database [29], [30] is a collection of 48 half-hour excerpts of two-channel ambulatory ECG recordings from 47 subjects studied by the BIH Arrhythmia Laboratory between 1975 and 1979. The recordings were digitized at 360 samples per second per channel with 11-bit resolution over a 10-mV range. Two or more cardiologists independently annotated each record, with disagreements resolved to obtain computer-readable reference annotations for each beat.

The National Metrology Institute of Germany (PTB) [31] has released a compilation of digitized ECGs for research, algorithmic benchmarking, and teaching purposes. The database contains 549 records from 290 subjects aged 17 to 87, 209 men, and 81 women. Each subject is represented by one to five records. Each record includes 15 simultaneously measured signals, including conventional 12 leads and 3 Frank lead ECGs. The signals are digitized at 1000 samples per second, with 16 bit resolution over a range of ± 16.384 mV. On special request, recordings may be available at sampling rates up to 10 KHz.

This dataset is composed of the abovementioned datasets, and all the samples are adjusted to a fixed dimension of 188 by [32], ensuring they are cropped, down-sampled, and padded with zeros if needed. These datasets have been used to explore heartbeat timeseries classification with a transformer model. The signals represent ECG heartbeat shapes for normal arrhythmia and myocardial infarction cases. The sample count is 109446, with 5 categories and a sampling frequency of 125Hz. The data was divided into train and test parts, with the train sample being 87554 and the test sample being 21892.

TABLE I. presents a concise overview of five distinct beat categories based on the guidelines set by the Association for the Advancement of Medical Instrumentation (AAMI) EC57 standard [28]. TABLE I indicates that the data is imbalanced, which poses a significant challenge where one or more classes are disproportionately represented compared to others. Moreover, we normalized the input data by removing the mean and scaling to unit variance (Zero-mean normalization). This normalization is calculated as $x\_normalize = (x - \mu)/\sigma$. Here,

$\mu$ and $\sigma$ represents the mean standard deviation of the input data $x$, respectively.

TABLE I. Details of dataset

| Category | Annotations | Instance Numbers Train/Test |
|---|---|---|
| N | Normal<br>Left/Right bundle branch block(LBBB or RBBB)<br>Atrial escape<br>Nodal escape<br>S Atrial premature(APB) | 72471/18118 |
| S | Atrial premature(APB)<br>Aberrant atrial premature<br>Nodal premature<br>Supra-ventricular premature | 2223/556 |
| V | Premature ventricular contraction(PVC)<br>Ventricular escape | 5788/1448 |
| F | Fusion of ventricular and normal | 641/162 |
| Q | Paced<br>Fusion of paced and normal<br>Unclassifiable | 6431/1608 |
| Total | | 87554/21892 |

## IV. Experimental Results and Discussions

The experiments were conducted using Python on a powerful system equipped with an Intel (R) Core (TM) 13900KF CPU @ 3.0GHz, 64 GB RAM, and a 16GB NVIDIA GeForce RTX 4080. The system was running on Windows 11. The models were implemented using Transformers, Scikit-Learn, and Keras on TensorFlow library. The Adam optimizer trained the model for 100 epochs with 32 batches and 1e-4 learning. We used Sparse Categorical Cross-entropy loss function. Later, SoftMax classifiers were used. TABLE II. provides details on model variants, while TABLE III. presents the training hyper-parameters used for model training. In this study, the model was trained from scratch, exclusively utilizing architectural design without the incorporation of pre-trained weights or prior knowledge from external datasets.

TABLE II. Details of model variants

| Parameter name | Values |
|---|---|
| Layers | 43 |
| Head size | 16 |
| Heads | 8 |
| Encoder Layers | 4 |
| MLP Unit | [128,64] |
| Input Dimension | $1 \times 177$ |
| Output Dimension | #classes (5) |
| Params | 36,301 |

TABLE III. Taining hyper-parameters

| Parameter names | | Values |
|---|---|---|
| Epoch | | 100 |
| Batch size | | 32 |
| Adam Optimizer | learning rate | 1e-4 |
| | $\beta_1$ | 0.9 |
| | $\beta_2$ | 0.999 |
| | $\epsilon$ | 1e-07 |
| Regularization | Dropout probability | 0.15 |
| loss function | | Sparse Categorical Cross entropy |
| classifiers | | SoftMax |

The model checkpoint is set to save only the best solution found during training based on the validation's loss function evaluation. This is achieved by monitoring the metric on a validation set during training and saving the model checkpoint only when the metric improves.

We have presented the classification performance of the proposed method on the MIT-BIH and PTB datasets. There are four metrics used to assess the effectiveness of a classification model: accuracy, precision, F-score, and recall (sensitivity). TABLE IV. reports classification metrics such as precision, recall, F1-score, and support for different heartbeat categories, along with overall accuracy metrics (accuracy, macro average, and weighted average) achieved by the ECGFORMER model.

The ECGFORMER model has demonstrated impressive classification performance across various heartbeat categories. It achieved high precision (0.98) for category N, indicating 98% of predicted instances were correct, while lower precision (0.88) was achieved for category F, indicating 88% of predicted instances were correct. The model also achieved perfect recall for category N, indicating the model correctly identified all instances, while category S had a lower recall (0.54), indicating 54% of actual instances were correctly identified. The F1-score, the harmonic mean of precision and recall, offers a balanced measure between the two metrics, considering both false positives and false negatives. For example, category Q had a high F1-score (0.99), indicating a balanced performance, while category F had a lower F1-score (0.66), suggesting a less balanced performance. The table presents classification results for a model with an overall accuracy of 98% across all categories. This metric represents the proportion of correctly predicted instances among the total instances in the dataset. The macro-average metric represents the unweighted average of precision, recall, and F1-score across all categories, giving equal importance to each category regardless of the number of instances. macro-average precision, recall, and F1-score are 0.95, 0.80, and 0.86, respectively. The weighted average metric considers the support (number of instances) for each category, giving more weight to categories with more instances. Balanced performance across categories is shown by the weighted-average precision, recall, and F1-score of 0.98.

TABLE IV.     CLASSIFICATION RESULTS ACHIEVED BY THE ECGFORMER

| Category | Precision | recall | f1-score | Support |
|---|---|---|---|---|
| N | 0.98 | 1 | 0.99 | 18118 |
| S | 0.97 | 0.58 | 0.73 | 556 |
| V | 0.95 | 0.92 | 0.93 | 1448 |
| F | 0.88 | 0.54 | 0.66 | 162 |
| Q | 0.99 | 0.97 | 0.98 | 1608 |
| | | | | |
| accuracy | - | - | 0.98 | 21892 |
| macro avg | 0.95 | 0.80 | 0.86 | 21892 |
| weighted avg | 0.98 | 0.98 | 0.97 | 21892 |

Accuracy metric in imbalanced datasets can be misleading, as accuracy may favor the majority class. The F1 score, a harmonic mean of precision and recall, is more informative in imbalanced datasets, considering both false positives and false negatives.

The curve of the loss function computed for both the training and validation datasets at each epoch throughout the model training process is illustrated in Fig. 3. Moreover, the accuracy of the model at each epoch on the train and validation datasets is depicted in the curve in Fig. 4.

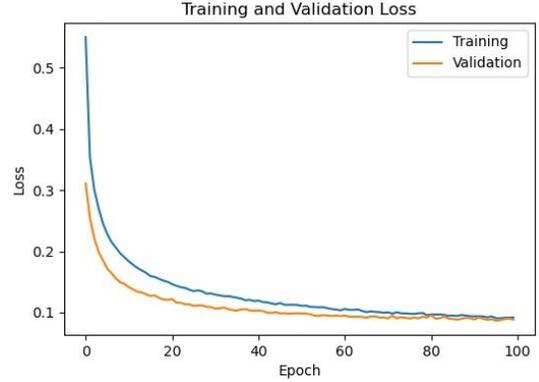

Fig. 3. Loss Curves: The plot illustrates the progression of the loss function computed on both the training and validation datasets across each epoch during the model training process.

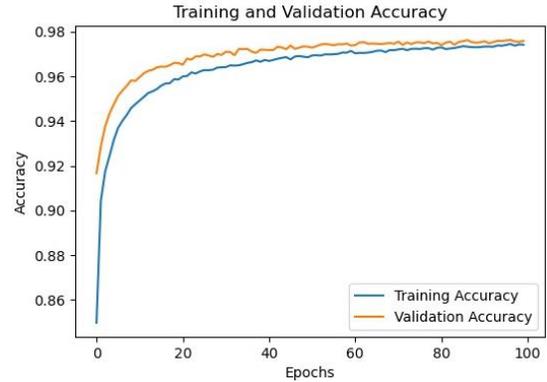

Fig. 4. Accuracy Curves: The depicted curve showcases the model's accuracy on the training and validation datasets at each epoch throughout the model training.

In Fig. 5, the classification confusion matrix for heartbeats on the test set illustrates the distribution of predicted instances for each class, providing insights into the model's performance and accuracy in classifying different heartbeat categories.

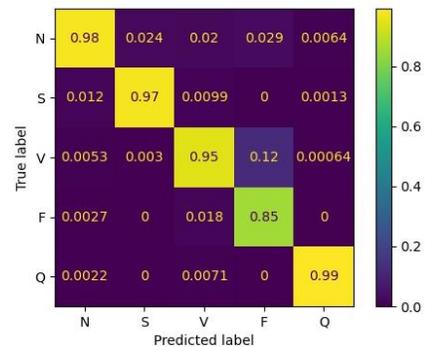

Fig. 5. Classification confusion matrix for heartbeats on the test set. The confusion matrix is normalized by the number of predicted instances for each class.

The researchers utilized a range of techniques to automatically classify heartbeat arrhythmias based on ECG signals obtained from the identical public dataset (MIT-BIH or PTB), as detailed in TABLE V.

TABLE V.   Comparison of classification outcomes for heartbeats

| Method | Dataset | Approach | Accuracy |
|---|---|---|---|
| [32] | MIT-BIH and PTB | Deep residual CNN | 93.4% |
| [33] | MIT-BIH | Augmentation-CNN | 94.03% |
| [34] | MIT-BIH | CNN-LSTM | 94.2 |
| ECGformer | MIT-BIH and PTB | Transformer-Attention | **98%** |

The table presents a comparative analysis of various methods for classifying ECG data from various datasets, revealing their reported accuracies. The Deep Residual CNN method achieved 93.4% accuracy, followed by Augmentation-CNN at 94.03% and CNN-LSTM at 94.2%. The ECGformer approach, a Transformer-based architecture with attention mechanisms, achieved the highest accuracy of 97%. The table shows the performance of different approaches and architectures in classifying ECG signals, with the ECGformer approach achieving the highest reported accuracy of 97% among the listed methods.

## V. Conclusion

This study focused on developing an effective framework, ECGformer, leveraging a ttransformer-based architecture with attention mechanisms for accurately classifying various heartbeat categories within electrocardiogram (ECG) data. The evaluation of the ECGFORMER model on the MIT-BIH and PTB datasets highlighted its exceptional performance in arrhythmia classification. The numerical values for the macro-average precision, recall, and F1-score are 0.95, 0.80, and 0.86, respectively. The numerical values for the weighted-average precision, recall, and F1-score are 0.98. Comparative analysis against other methods revealed the ECGFORMER's superiority, with an accuracy of 98%, surpassing the performance of Deep Residual CNN (93.4%), Augmentation-CNN (94.03%), and CNN-LSTM (94.2%). The study demonstrates the effectiveness of Transformer-based architectures, specifically ECGformer, in detecting precise arrhythmias in ECG signals. The results suggest its potential for improved cardiac healthcare diagnostics and clinical decision-making. The ECGformer model represents a significant advancement in deep learning techniques for cardiac health monitoring and diagnosis.

## Acknowledgment

The data used in the preparation of this article is composed of two collections of heartbeat signals derived from the MIT-BIH Arrhythmia Dataset and the PTB Diagnostic ECG Database. All the samples were pre-processed by https://arxiv.org/abs/1805.00794 and available on Kaggle.